\documentclass{PoS}
\usepackage{mathptmx}
\usepackage{calrsfs}
\usepackage{pdfpages}
\DeclareMathAlphabet{\mathcal}{OMS}{zplm}{m}{n}

\let\OLDthebibliography\thebibliography
\renewcommand\thebibliography[1]{
  \OLDthebibliography{#1}
  \setlength{\parskip}{0pt}
  \setlength{\itemsep}{-3pt}
\footnotesize
}

\def\vc#1{\mbox{\boldmath $#1$}}

\title{Enhancements and suppressions of CP violating effect in the nucleons, nuclei, and atoms}

\ShortTitle{Enhancements and suppressions of CP violating effect in the nucleons, nuclei, and atoms}

\author{\speaker{Nodoka~Yamanaka}\thanks{Supported by JSPS Postdoctoral Fellowships for Research Abroad.}\\
        IPNO, CNRS-IN2P3, Univ. Paris-Sud, Universit\'e Paris-Saclay, 
91406 Orsay Cedex, France\\
        E-mail: \email{yamanaka@ipno.in2p3.fr}
}

\abstract{
The electric dipole moment (EDM) is a very sensitive probe of CP violation beyond the standard model, and it as been measured in many systems such as atoms, neutrons, etc.
The EDM of composite systems may be sensitive to several elementary level CP violating processes, but the theoretical evaluations of the CP violation at different physical (atomic, nuclear, hadronic, elementary) hierarchies are required to unveil them.
In this context, we are particularly interested in which CP violating processes are enhanced in a given system, or vice versa.
In this proceedings contribution, we will give an overview of the enhancement and suppression of CP violation in processes contributing to the EDMs of composite systems.
}

\FullConference{23rd International Spin Physics Symposium - SPIN2018 -\\
		10-14 September, 2018\\
		Ferrara, Italy}

\begin{document}

\section{Introduction}

The CP violation of the standard model (SM) is not sufficient to explain the matter/antimatter asymmetry of the Universe \cite{sakharov,farrar,huet}.
An attractive approach to search for new physics beyond the SM with large CP violation is the measurement of the {\it electric dipole moment} (EDM) \cite{hereview,Bernreuther,barrreview,khriplovichbook,ginges,pospelovreview,fukuyama,engelreview,yamanakabook,devriesreview,yamanakanuclearedmreview,atomicedmreview,chuppreview}.
It is now possible to experimentally measure the EDM of many systems (neutron \cite{baker}, atoms \cite{parker,graner}, muon \cite{muong-2}, electron in molecules \cite{hudson,Andreev:2018ayy}, etc).
There are also much R \& D with new experimental techniques to explore other systems (protons \cite{Anastassopoulos}, nuclei \cite{jedi,jedi2}, heavy hadrons \cite{Baryshevsky}, $\tau$ leptons \cite{xinchen,koksal,koksal2,Koksal:2018vtt}, electrons in new systems \cite{inertgasmatrix,Kozyryev:2017cwq},...).
We note that one of the main topic discussed in the Session "Fundamental Symmetries and Spin Physics BSM" of this symposium is the general relativity effect to the EDM measurement using storage rings.
This topic was investigated in many previous works \cite{Kobach:2016kvn,Silenko:2006er,Obukhov:2016vvk,Obukhov:2016yvw,Orlov,Laszlo:2018llb} and an excellent derivation of this effect was also presented in the same session \cite{Laszlo:2019cyu}, so we will not expand the physics in this direction.
Here we will rather concentrate on how to extract new theories from the EDM experimental data.
For that, we must know how the elementary level CP violation and the EDM of the systems seen above are related in theory.
We are particularly interested in systems where enhancements occurs, or those with specific experimental sensitivity to a definite class of CP violating theories.
In this proceedings contribution, we will give a (disorganized) review of the mechanisms of enhancement and suppression of hadronic, nuclear and atomic level CP violation contributing to the EDM of various systems currently studied in experiments.

\section{Electric dipole moments of elementary and composite systems\label{sec:elementary_edm}}

In the relativistic field theory, the interaction of the EDM of an elementary fermion $f$ with the external electric field is expressed as
\begin{equation}
{\cal L}_{\rm EDM}
=
-\frac{i}{2} d_f \bar f \sigma_{\mu \nu} F^{\mu \nu} \gamma_5 f
.
\label{eq:edm}
\end{equation}
By taking the nonrelativistic limit, the above lagrangian reduces to $H_{\rm EDM} = -d_f \vc{\sigma}\cdot {\mathbf E}$, where $\vc{\sigma}$ is the spin of $f$.
This effective interaction is generated in many candidates of new theories at one- or two-loop levels, such as in supersymmetric theories \cite{ellismssmedm,chang,splitsusy1,rpv3}, extended Higgs models \cite{barr-zee,Chen:2017com,Egana-Ugrinovic:2018fpy}, etc.
The SM contribution appears at the three-loop level or beyond \cite{Czarnecki:1997bu,Pospelov:2013sca}, and is thus extremely small.
This tiny background makes the EDM to be a probe very sensitive to the CP violation.

For the case of composite particles, the evaluation of the EDM proceeds with the calculation of 
\begin{equation}
{\mathbf d}_\Psi
=
\sum_{i} \langle \Psi | Q_i e {\mathbf r}_i | \Psi \rangle
,
\end{equation}
with $Q_i e$ and ${\mathbf r}_i$ the charge and the coordinate (in the center of mass frame) of the $i$th constituent of the system $\Psi$, respectively.
Here the EDM is not only generated by that of elementary particles, but also by other CP-odd interactions which may polarize the entire system.
The leading ones with low mass dimensions are the followings:
\begin{eqnarray}
{\cal L}_{\rm cEDM}
&=&
-\frac{i}{2} d_f^c \bar f \sigma_{\mu \nu} G^{\mu \nu}_a t_a \gamma_5 f
\ \ \ \ \ (\mbox{chromo-EDM}),
\label{eq:cedm}
\\
{\cal L}_{\rm w}
&=&
\frac{1}{3!} w 
f^{abc} \epsilon^{\alpha \beta \gamma \delta} G^a_{\mu \alpha } G_{\beta \gamma}^b G_{\delta}^{\mu,c}
\ \ \ \ \ (\mbox{Weinberg operator}),
\label{eq:weinbergop}
\\
{\cal L}_{\rm ff'}
&=&
C_{i , ff'} \bar f \Gamma_i f \, \bar f \Gamma'_i f
\ \ \ \ \ (\mbox{4-fermion interaction})
.
\label{eq:4-fermion}
\end{eqnarray}
Each of them may be generated at low orders of perturbation in many candidates of new physics, and has specific sensitivity to them.

In neutral atoms or molecules, an important suppression of the CP violating effect occurs.
Schiff showed that the EDM of pointlike constituents in nonrelativistic and charge neutral systems is completely screened \cite{schiff}.
There are however several effects which escape from Schiff's theorem.
These are (1) the relativistic constituents, in particular the electron in heavy atoms or molecules, (2) the CP violating interactions among constituents, and (3) the nuclear finite size effect, given by the nuclear Schiff moment
\begin{equation}
{\mathbf S}
\equiv
\sum_{N=1}^A
\Biggl[
\frac{1}{6}
\biggl(
\frac{1}{10}r_N^2 
-\langle r^2 \rangle_{\rm ch}
\biggr)
{\mathbf d}_N
+\frac{1}{5}
({\mathbf r}_N \cdot {\mathbf d}_N ) {\mathbf r}_N
\Biggr]
+
\frac{e}{10}
\sum_{i=1}^Z
\Bigl(
r_i^2 - \frac{5}{3} \langle r^2 \rangle_{\rm ch}
\Bigr)
{\mathbf r}_i
,
\end{equation}
where ${\mathbf d}_N$ is the intrinsic nucleon EDM.
The indices $i$ and $N$ run over all protons and nucleons of the nucleus with charge radius $\langle r^2 \rangle_{\rm ch}$, respectively.
To extract the new physics from atomic or molecular experiments, the quantifications of all of the above physics, at the elementary, hadronic, nuclear, and atomic levels, are mandatory.
On that occasion, knowing enhancing or suppressing mechanisms may help us to achieve this goal.

\section{Enhancement and suppression of the EDM in composite systems}

\subsection{Relativistic enhancement of the electrion EDM}

In heavy paramagnetic atoms or molecules, the internal electric field is huge and electrons become relativistic.
The system is then polarized by the second order perturbation, as \cite{sandars}
\begin{equation}
d_a
=
\sum_{n} \sum_{i,j}^Z
\frac{\langle \Psi_0 | -e z_i | \Psi_n \rangle \langle \Psi_n | d_e (1-\beta_j) \vc{\sigma}_j \cdot {\mathbf E}_i | \Psi_0 \rangle }{E_n -E_0}
+{\rm c.c.}
,
\end{equation}
where $\beta $ is the zeroth component of the Dirac gamma matrix, so that $(1-\beta)$ projects out the nonrelativistic contribution.
The relativistic effect grows as $Z^3$ and amplifies the electron EDM in heavy atoms or molecules.
For instance, the enhancement factor of $d_e$ is more than 500 times for the Tl atom \cite{Dzuba:2009mw,Porsev:2012zx}, and about 900 times for Fr \cite{flambaumfr,mukherjeefr}.

\begin{figure}[hbt]
\begin{center}
\includegraphics[clip,width=.76\columnwidth]{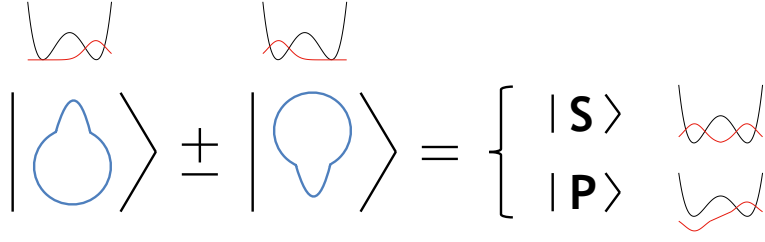}
\caption{
Schematic explanation of the parity doubling in octupole deformed systems.
The physical states are the superposition of the two orientations of the octupole deformation, which can be regarded as the two almost degenerate localized states of the double well potential.
The red lines are describing the wave function in the double well potential drawn with solid black curves.
The resulting S- and P-wave states on the right-hand side of the equation have close energy levels but are not completely degenerate due to the tunneling between the wells.
}
\label{fig:parity_doubling}
\end{center}
\end{figure}

This effect becomes much more important in polar molecules due to the very large effective electric field.
It reaches 23 GV/cm for YbF \cite{abe,sunaga}, and $-$79 GV/cm for ThO \cite{Denis,Skripnikov}.
This enhancement of the electron EDM is explained by the parity doubling \cite{Labzovskii,Sushkovmolecule,Kozlov:1994zz,Kozlov:1995xz,Chubukov}.
Indeed, the two orientations of the polar molecule can be regarded as the two localized states of the double well potential problem, for which the parity even and odd superpositions of the latter have very close energy levels (see Fig. \ref{fig:parity_doubling}).
These two localized states just correspond to the two orientations of the polar molecule.

The electron EDM also contributes to the diamagnetic atoms via the hyperfine interaction, but this effect is much more suppressed than the paramagnetic one \cite{ginges,flambaum,Fleig:2018bsf}.

\subsection{Renormalization group evolution of quark-gluon level CP violation\label{sec:RGE}}

The elementary level CP-odd interactions of Eqs. (\ref{eq:edm}), (\ref{eq:cedm}), (\ref{eq:weinbergop}), and (\ref{eq:4-fermion}) are generated at the energy scale of the new physics (TeV or beyond), after integrations of heavy particles.
When we consider them at a lower energy scale (for instance at the hadronic scale), however, they receive contribution from the integration of other particles of the SM (threshold correction) and from the mixing between each other due to the radiative corrections.
The variation of their Wilson coefficients can be calculated using the perturbative renormalization group evolution.
This effect is especially important at low energy scale for CP-odd interactions involving quarks and gluons.
For instance, the leading perturbative QCD evolution of the isoscalar quark mass $m_l \equiv \frac{m_u + m_d}2$, isoscalar quark EDM $d_q^{(0)}$, isoscalar chromo-EDM $d_q^{c,(0)}$, and the Weinberg operator from the TeV scale ($M_{\rm NP}=1$ TeV) to the hadronic scale ($\mu_{\rm had} = 1$ GeV) yields \cite{Degrassi:2005zd}
\begin{eqnarray}
m_l (M_{\rm NP})
&=&
0.5 m_l  (\mu_{\rm had})
,
\\
d_q^{(0)} (M_{\rm NP})
&=&
0.8 d_q^{(0)} (\mu_{\rm had})
,
\\
d_q^{c , (0)} (M_{\rm NP})
&=&
0.9 d_q^{c ,(0)} (\mu_{\rm had})
-0.8 d_q^{(0)} (\mu_{\rm had}) / e (\mu_{\rm had})
,
\\
w (M_{\rm NP})
&=&
0.16 w (\mu_{\rm had})
-0.14 d_q^{c ,(0)} (\mu_{\rm had}) / m_l (\mu_{\rm had})
+0.08 d_q^{(0)} (\mu_{\rm had}) /[m_l (\mu_{\rm had}) e (\mu_{\rm had})]
.
\ \ \ \ \ \ 
\end{eqnarray}
Here we neglected the effect of four-quark operators \cite{Hisano:2012cc,Buchalla:1995vs} as well as the electroweak threshold corrections due to heavy particles \cite{Dekens:2013zca}.
From the above result, we see that the $d_q^{(0)}$, $d_q^{c,(0)}$, and $w$ are suppressed due to the QCD corrections.
This trend can actually be understood by the suppression of the quark/gluon spin due to the superposition of the successive gluon emission/absorption processes \cite{Yamanaka:2013zoa}.
On the other hand, the current quark mass is suppressed during the evolution down to the low energy scale, which means that the Wilson coefficients of the $\bar q q$ operator increased (recall that $m_q \bar qq$ is invariant under renormalization).
This enhancement can be explained by the elongation of the quark world line, whose length contributes constructively to the quark scalar density \cite{Yamanaka:2014lva}.

\subsection{CP-odd electron-nucleon interactions}

The CP-odd $e-N$ interaction can only be studied in atomic EDM and molecular beam experiments, but it is interesting since it has a specific sensitivity to several classes of models such as the Higgs doublet models \cite{Barr:1991yx}, supersymmetric models wit large $\tan \beta$ \cite{Lebedev:2002ne} or with R-parity violation \cite{herczege-n,rpvlinearprogramming}, leptoquark models \cite{herczegleptoquark,Fuyuto:2018scm}, etc.
Other important points are that its contribution is enhanced by the many-body physics and that it can be evaluated with a reasonable accuracy.

The leading CP-odd $e-N$ interaction is given by
\begin{equation}
{\cal L}_{eN}
=
-\frac{G_F}{\sqrt{2}} \sum_{N=p,n} \left[
C_N^{\rm SP} \bar NN \, \bar e i \gamma_5 e
+C_N^{\rm PS} \bar Ni\gamma_5 N \, \bar e e 
-\frac{1}{2}C_N^{\rm T} \epsilon^{\mu \nu \rho \sigma} \bar N \sigma_{\mu \nu} N \, \bar e \sigma_{\rho \sigma} e 
\right] 
.
\ \ \ \ \ 
\label{eq:pcpve-nint}
\end{equation}
The couplings ($C_N^{\rm SP, PS, T}$) are simply obtained by multiplying the nucleon matrix elements by the CP-odd electron-quark/gluon couplings with the same Lorentz and flavor structures, which were evolved down to the hadronic scale with the renormalization procedure of Section \ref{sec:RGE}.

\begin{figure}[hbt]
\begin{center}
\includegraphics[width=.46\columnwidth]{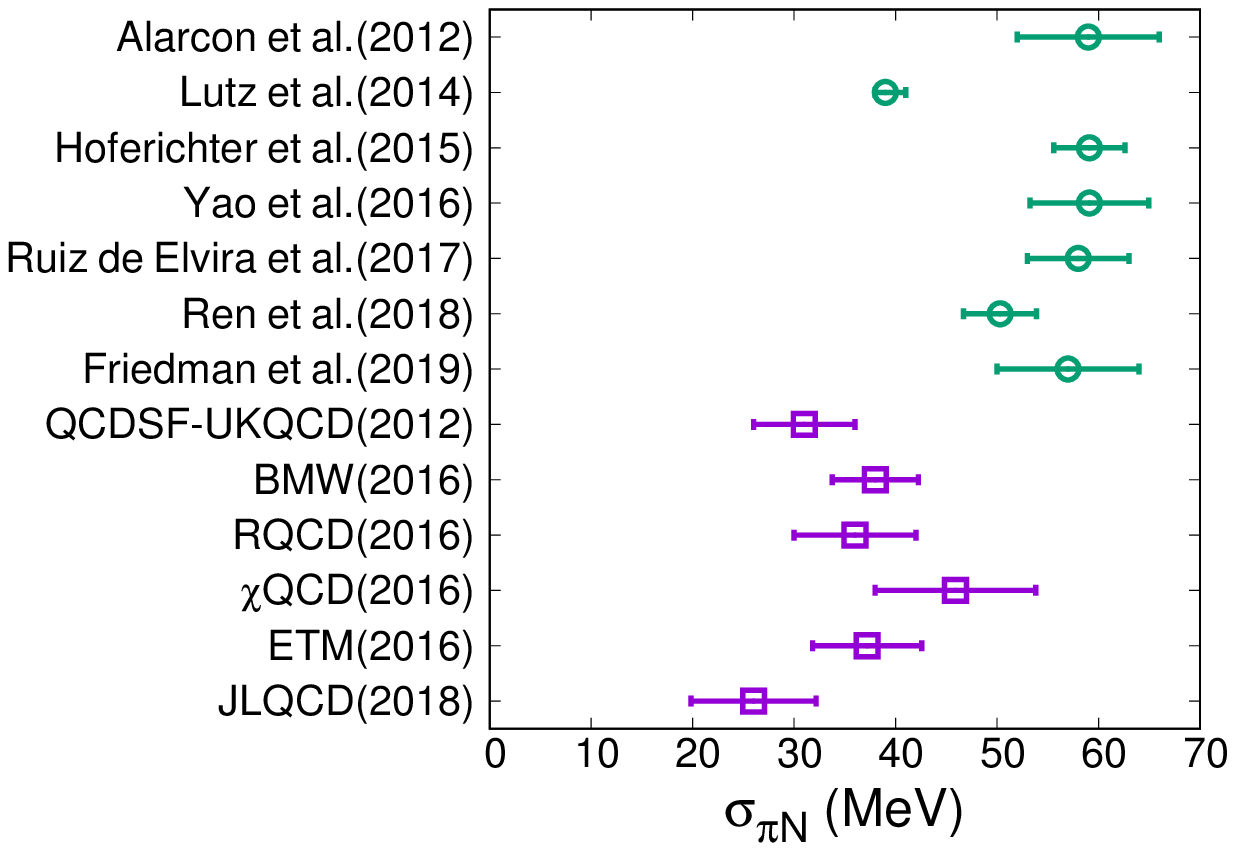}
\includegraphics[width=.46\columnwidth]{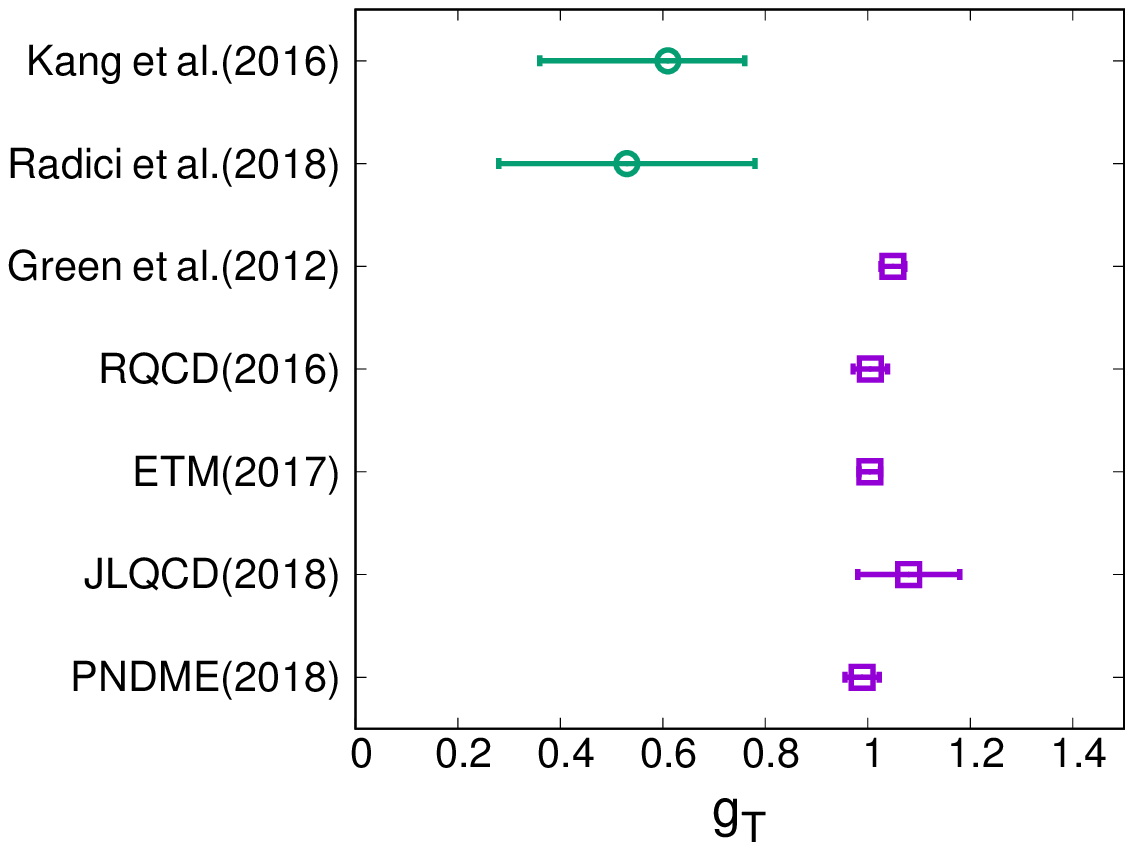}
\caption{
Summary of phenomenological and lattice QCD results of the calculations of the nucleon sigma term $\sigma_{\pi N} \equiv m_l \langle N| \bar uu + \bar dd|N\rangle $ \cite{alarcon,Lutz:2014oxa,Hoferichter,yao,deelvira,Ling:2017jyz,Friedman:2019zhc,chiqcdsigmaterm,rqcdsigmaterm,bmwsigmaterm,etmsigmaterm,Yamanaka:2018uud} and the isovector tensor charge $g_T \equiv \langle p| \bar ui \sigma_{03} \gamma_5 u -\bar d i \sigma_{03} \gamma_5 d |p\rangle $ \cite{kang,radici2,green,rqcdisovector,etm2017,Yamanaka:2018uud,pndmetensor}.
}
\label{fig:sigma_term_tensor_charge}
\end{center}
\end{figure}

The nucleon matrix elements required in this step are the scalar $\langle N| \bar qq|N\rangle $, pseudoscalar\\ $\langle N| \bar qi\gamma_5 q|N\rangle $, and tensor ones $\delta q \equiv \langle p| \bar qi \sigma_{03} \gamma_5 q |p\rangle $.
The nucleon scalar and tensor charges have been extensively studied, phenomenologically and on lattice.
The current results are summarized in Fig. \ref{fig:sigma_term_tensor_charge}.
Although having some systematic deviations between approaches, they are determined with good precision.
Here we note that the isoscalar quark scalar density $\langle N| \bar uu + \bar dd|N\rangle $ is close to ten, which is larger than the nonrelativistic prediction $\langle N| \bar uu +\bar dd|N\rangle = 3$, whereas the tensor charge is at most one, being smaller than the nonrelativistic value 5/3.
If we interpret this in terms of the superposition of gluon radiation, this mechanism is even more pronounced in the nonperturbative generation of the nucleon charges \cite{Yamanaka:2013zoa,Yamanaka:2014lva}.

For the calculation of the PS type CP-odd $e-N$ interaction [second term of Eq. (\ref{eq:pcpve-nint})], the nucleon pseudoscalar density $\langle N| \bar qi\gamma_5q|N\rangle $ is required.
It can be calculated by relating it to the nucleon axial charge via the anomalous Ward identity \cite{herczege-n,Yamanaka:2012zy}.
At the leading order of chiral perturbation, we have $\langle p | \bar u i \gamma_5 u | p \rangle = 180$, $\langle p | \bar d i \gamma_5 d | p \rangle = -170$ \cite{atomicedmreview}.
These large values are due to the pion pole effect which is inversely proportional to the light quark mass.
A remarkable point is that this enhancement compensates the nonrelativistic suppression of the nucleon pseudoscalar density ($\propto {\mathbf p} \cdot \vc{\sigma} $) at the atomic level.

At the nuclear level, the effect of $C_N^{\rm SP}$ is enhanced by the nucleon number.
In paramagnetic systems, it is moreover amplified by the relativistic effect \cite{dzubarelation,Jung:2013mg}, whereas it is suppressed for the diamagnetic ones due to the closed electron shell, like for the electron EDM \cite{ginges,flambaum,Fleig:2018bsf}.
On the other hand, $C_N^{\rm PS}$ and $C_N^{\rm T}$ contribute to the atomic EDM through the nuclear spin, and the value of its matrix element $\langle \psi | \vc{\sigma}_N | \psi \rangle$ is required.
For the cases of $^{129}$Xe and $^{199}$Hg, the results of the sophisticated shell model calculations $\langle ^{129}{\rm Xe} | \vc{\sigma}_n | ^{129}{\rm Xe} \rangle \approx 0.2$ \cite{yoshinaga1,yoshinaga2} and $\langle ^{199}{\rm Hg} | \vc{\sigma}_n | ^{199}{\rm Hg} \rangle \approx -0.4$ \cite{Yanase:2018qqq} are available, with 30\% accuracy.
The nuclear spin matrix elements are in general smaller than one due to the pairing of nucleons, and are further suppressed by the mixing of the nucleon spin with the orbital angular momentum.
At the atomic level, the atomic polarization due to the tensor type CP-odd $e-N$ interaction grows with $Z^2$ and becomes important for heavy atoms.
Modern many-body calculations are giving the linear coefficients relating $C_N^{\rm T}$ to the atomic EDM with typical accuracy of a few percent level \cite{Dzuba:2009kn,sahoo2,Fleig:2018etq}.
As we saw above, the PS type CP-odd $e-N$ interaction is nonrelativistically suppressed at the atomic level, but the large nucleon pseudoscalar matrix elements and the very high precision of the EDM of diamagnetic atoms (especially the $^{199}$Hg atom \cite{graner}) make it interesting in constraining the new physics.

\subsection{Chiral enhancement of the nucleon EDM}

\begin{figure}[hbt]
\begin{center}
\includegraphics[width=.36\columnwidth]{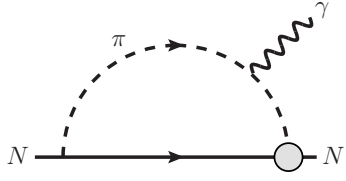}
\caption{
Leading contribution to the nucleon EDM in chiral perturbation.
The grey blob represents the CP-odd pion-nucleon interaction.
}
\label{fig:nucleon_EDM}
\end{center}
\end{figure}

The nucleon EDM is generated by several CP-odd quark-gluon level processes, and all of them require nonperturbative calculations.
The only one which could fully be obtained from first principle is the quark EDM contribution ($d_p (d_q) = \sum_q \delta q \, d_q $, see previous section for the definition of the tensor charge $\delta q$), from lattice QCD (see also right panel of Fig. \ref{fig:sigma_term_tensor_charge}).
The most interesting cases are however the quark chromo-EDM and the CP-odd four-quark interactions, which are generated by popular models such as the supersymmetric SM \cite{pospelovreview,ibrahimmssmedm,chang,mssmreloaded,mssmrainbow} or the left-right symmetric models \cite{Xu:2009nt,deVries:2012ab,Maiezza:2014ala,Dekens:2014jka}.
Although no first principle calculations are available, it is possible to expand the unknown CP-odd couplings in terms of symmetry breaking parameters using the chiral effective field theory (EFT) and to quantify them by finding the leading contribution.
The leading order nucleon EDM is given by \cite{engelreview,devriesreview,yamanakanuclearedmreview,chuppreview,Crewther:1979pi,ottnad,Mereghetti:2010kp,guo,Fuyuto:2012yf,deVries:2015una}
\begin{equation}
d_N 
=
\bar d_N 
+\tau_3
\frac{e g_A \bar g^{0} }{2\pi^2 f_\pi}
\ln \frac{\Lambda^2}{m_\pi^2}
,
\label{eq:nucleon_EDM}
\end{equation}
where $\bar g^{0}$ is the isoscalar CP-odd pion-nucleon coupling, $\Lambda \sim 1$ GeV is the hadronic cutoff, and $\tau$ the isospin Pauli matrix.
The counterterm $\bar d_N $ comprises all short distance effect finer than $1/\Lambda$, including the contribution from the quark EDM, as seen above.
The quark chromo-EDM and the CP-odd four-quark interactions are chiral symmetry breaking, so their contribution to the nucleon EDM is in principle suppressed by one power of $m_\pi^2$.
However, these interactions generate the CP-odd pion-nucleon coupling, which intermediately contributes to the nucleon EDM (see Fig. \ref{fig:nucleon_EDM}), through the pion pole and the vacuum alignment, which bring a factor of $m_\pi^{-2}$, thus canceling the above suppression.
We also note that this one-loop level contribution is enhanced by $\ln (\Lambda / m_\pi )$ \cite{Crewther:1979pi}.
The effect of the quark chromo-EDM and the CP-odd four-quark interactions are therefore not small at the nucleon level.

The Weinberg operator [see Eq. (\ref{eq:weinbergop})] \cite{weinbergop} also directly contributes to the nucleon EDM since it is a purely gluonic interaction without chiral symmetry breaking.
The relation between them is however very obscure, and no quantitative results are currently available.
The Weinberg operator is sensitive to many interesting models such as the extended Higgs models \cite{weinbergop} or vectorlike quark models \cite{Choi:2016hro}, so more detailed quantifications are definitely required in the future.

\subsection{Nuclear EDM and nuclear Schiff moment from CP-odd nuclear force}

The combination of the CP-odd pion-nucleon interaction with the CP-even one generates the CP-odd nuclear force, which polarizes the nuclear systems and generates CP-odd moments.
At the leading order in chiral EFT, we have \cite{pvcpvhamiltonian1,pvcpvhamiltonian2,pvcpvhamiltonian3}
\begin{eqnarray}
H_{P\hspace{-.5em}/\, T\hspace{-.5em}/\, }^\pi
& = &
-\frac{m_\pi }{2 m_N} 
\bigg\{ 
\bar{G}_{\pi}^{(0)}\,{\vc{\tau}}_{1}\cdot {\vc{\tau}}_{2}\, (\vc{\sigma}_1 - \vc{\sigma}_2) 
+ 
\bar{G}_{\pi}^{(1)}\,
( \tau_{1}^{z}\, {\vc{\sigma}}_{1} -\tau_{2}^{z}\,{\vc{\sigma}}_{2} )
\bigg\}
\cdot \frac{{\mathbf r}_1 - {\mathbf r}_2}{|{\mathbf r}_1 - {\mathbf r}_2|} 
\frac{e^{-m_\pi r }}{4\pi r} \left( 1+ \frac{1}{m_\pi r} \right)
,
\ \ \ \ \ \ \ 
\label{eq:CPVhamiltonian}
\end{eqnarray}
where the subscripts 1 and 2 label the two interacting nucleons.
We note that the above CP-odd nuclear potential is spin dependent, so only nuclei with open spin shell can have CP-odd moments.
In chiral perturbation, the isoscalar coupling $\bar G_\pi^{(0)}$ is related to $\bar g_0$ of Eq. (\ref{eq:nucleon_EDM}) by $\bar G_\pi^{(0)} = \frac{g_A m_N}{f_\pi} \bar g_0$.

\begin{figure}[hbt]
\begin{center}
\includegraphics[clip,width=.98\columnwidth]{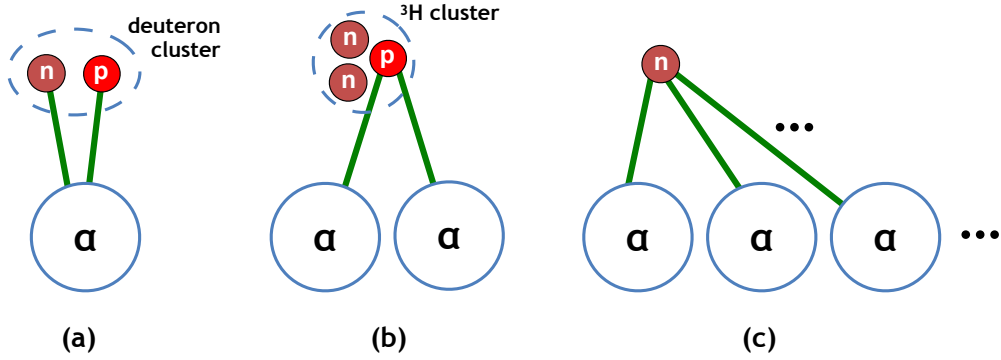}
\caption{
Examples of the counting rule of nuclear EDM for (a) $^6$Li, (b) $^{11}$B, and (c) heavy nuclei.
}
\label{fig:cluster_edm}
\end{center}
\end{figure}

We first consider the EDM of light nuclei, which is expected to be measured in experiments with very high accuracy \cite{Anastassopoulos,jedi,jedi2,bnl}.
We emphasize that the EDM of bare nuclei is not damped by Schiff's screening.
With the CP-odd nuclear potential (\ref{eq:CPVhamiltonian}), the EDMs of several light nuclei have been calculated \cite{yamanakanuclearedmreview,liu,bsaisou,yamanakanuclearedm,c13edm,Yamanaka:2018dwa}, and an approximate counting rule 
\begin{equation}
d_A
= 
(\mbox{cluster EDM})
+
N_{\alpha N} \times (\alpha -N \, \mbox{polarization}, \propto \bar G_\pi^{(1)})
,
\end{equation}
was derived, where $N_{\alpha N}$ counts the number of $\alpha - N$ subsystems without closed spin shell (some examples of the counting are given in Fig. \ref{fig:cluster_edm}).
Large $N_{\alpha N}$ may lead to an enhancement of the sensitivity to $\bar G_\pi^{(1)}$.
This fact can be explained by the scalar density dependence of the CP-odd nuclear force (\ref{eq:CPVhamiltonian}).
This cannot be applied to nuclei which have bad overlaps between opposite parity states, such as the $^{13}$C \cite{c13edm}.
From a naive extrapolation, the nuclear EDM becomes very large for heavier nuclei [see Fig. \ref{fig:cluster_edm} (c)].
However, it is known that the configuration mixing is relevant in heavy nuclear systems, and the destructive interference of angular momenta actually suppresses the nuclear EDM for them \cite{atomicedmreview,yoshinaga1,yoshinaga2}.

As we saw previously, the CP violation of heavy nuclei contributes to the atomic EDM through the nuclear Schiff moment.
It has been calculated for experimentally interesting nuclei in several approaches ($^{129}$Xe \cite{yoshinaga2,Dmitriev:2004fk,yoshinaga3,teruya}, $^{199}$Hg \cite{Ban:2010ea}, $^{225}$Ra \cite{Dobaczewski:2005hz}).
Like the nuclear EDM, the nuclear Schiff moment of $^{129}$Xe calculated with the core polarization effect is much more suppressed than the simple shell model estimation \cite{Dmitriev:2004fk}.
This suppression is explained by the destructive interference of angular momentum due to the configuration mixing.

The most interesting cases are the octupole deformed nuclei, such as $^{225}$Ra or $^{223}$Rn, for which the sensitivity of the Schiff moment to the CP-odd nuclear force may be enhanced by more than three orders of magnitude \cite{Dobaczewski:2005hz,Dobaczewski:2018nim}.
The enhancement is due to the parity doubling, just like the polar molecules (see Fig. \ref{fig:parity_doubling}).
The calculations of the Schiff moment of heavy nuclei are not accurate for some nuclei and the theoretical uncertainty may exceed 100\% \cite{Ban:2010ea}.
Further investigations using several different approaches are therefore still required for quantification.
The atomic level calculations of the Schiff moment contribution are however very accurate, and the error bars may be small as a few percents \cite{ginges,atomicedmreview,Dzuba:2009kn,singh1,Ramachandran,lathahg,radziute,singh2,sahoo,singh3,Sahoo:2017tlw}.

\subsection{Standard model contribution}

As we saw in Section \ref{sec:elementary_edm}, the SM contribution to the EDM of elementary fermions is tiny ($d_{u,d} \sim 10^{-35}e$ cm from three-loop level calculation \cite{Czarnecki:1997bu}, $d_e \sim 10^{-44}e$ cm from estimation based on the cancellation at the three-loop level \cite{Pospelov:2013sca,Pospelov:1991zt}).
For the case of composite system, however, the EDM may be larger due to the hadron level processes.
The CP phase of the CKM matrix \cite{ckm} becomes relevant when the matrix elements are combined so as to form the Jarlskog invariant \cite{Jarlskog:1985ht}.
At the hadron level, this is possible through two distinct $|\Delta S|=1$ interactions, with one generated by the tree level $W$ boson exchange diagram, and the other induced by the penguin diagram.
We also point that the latter contribution is enhanced by the renormalization group evolution by more than an order of magnitude \cite{Buchalla:1995vs,yamanakasmcpvnn}.
With these hadronic inputs, the nucleon EDM generated by the CKM matrix is $d_N \sim 10^{-32}e$ cm \cite{seng}.
At the nuclear level, the CP-odd nuclear couplings are generated with an estimated value $\bar G_\pi^{(0,1)} \sim 10^{-17}$ \cite{yamanakasmcpvnn}, which gives the leading contribution to the EDM of nuclei and diamagnetic atoms \cite{yamanakasmdeuteronedm,Lee:2018flm}.
For the paramagnetic systems, the most important effect is due the CP-odd $e-N$ interaction $C_N^{\rm SP} \sim 10^{-17}$ generated at the hadronic level \cite{Pospelov:2013sca}, which largely exceeds that of the electron EDM generated at the elementary level.

\section{Summary}

In this short review, we summarized the mechanisms of enhancement and suppression of the CP violation in composite systems.
The most notable effect is the amplification of the EDM of electron in polar molecules and that of the nuclear Schiff moment of octupole deformed nuclei, which are both due to the parity doubling effect.
The scalar density at the elementary, hadronic, and nuclear levels may also enhance the EDM of composite systems.
On the other hand, spin related quantities of the strong interacting sector are suppressed by the mixing with other angular momentum components.
Needless to say, the most important damping mechanism is Schiff's screening in nonrelativistic charge neutral bound states.
We have to note that systems in which the EDM is experimentally measurable are not numerous, and the information which can be extracted is limited.
It is then desirable to evaluate all leading order relations between the elementary CP violation (with effective interactions with low mass dimension) and the observables measurable with sufficient accuracy to disentangle the new physics, even if the works to be done are still numerous due to the system dependence of the sensitivity to elementary CP-odd processes. 
We definitely encourage physicists to continue the quantitative study of the EDM of existing experimental projects.

\bibliographystyle{Science}

\bibliography{yamanaka}

\end{document}